\documentclass[12pt,finalcls,onecolumn]{IEEEtran}
\usepackage{graphicx}
\usepackage{algorithm}
\usepackage{algorithmic}
\usepackage{bm}
\usepackage{amsmath,amssymb}
\usepackage{dsfont}

\def\u{{\bf u}}

\def\x{{\bf x}}
\def\y{{\bf y}}

\def\b{{\bf b}}

\def\e{{\bf e}}

\begin{document}

\sloppy

\title{Iterative Learning for Reference-Guided DNA Sequence Assembly from Short
Reads: Algorithms and Limits of Performance}


\author{Xiaohu Shen, Manohar Shamaiah, and Haris Vikalo
\thanks{H. Vikalo and X. Shen are with the Department of Electrical and Computer Engineering,
The University of Texas at Austin, USA.}
\thanks{M. Shamaiah is with Broadcom Inc., Bangalore, India.}
}

\date{January 26, 2014}

%



\maketitle

\begin{abstract}

Recent emergence of next-generation DNA sequencing technology has enabled
acquisition of genetic information at unprecedented scales.
In order to determine the genetic blueprint of an organism,
sequencing platforms typically employ so-called shotgun sequencing strategy to
oversample the target genome with a library of relatively short overlapping reads. The
order of nucleotides in the reads is determined by processing the acquired noisy signals
generated by the sequencing instrument. Assembly of a genome from potentially erroneous
short reads is a computationally daunting task even in the scenario where a reference
genome exists. Errors and gaps in the reference, and perfect repeat regions in the target,
further render the assembly challenging and cause inaccuracies. In this paper, we
formulate the reference-guided sequence assembly problem as the inference of the
genome sequence on a bipartite graph and solve it using a message-passing algorithm.
The proposed
algorithm can be interpreted as the well-known classical belief propagation scheme
under a certain prior. Unlike existing state-of-the-art methods,
the proposed algorithm combines the information provided by the reads
without needing to know reliability of the short reads (so-called quality scores).
Relation of the message-passing algorithm to a provably convergent power iteration
scheme is discussed. To
evaluate and benchmark the performance of the proposed technique, we find an
analytical expression for the probability of error of a genie-aided maximum a
posteriori (MAP) decision scheme. Results on both simulated and experimental
data demonstrate that the proposed message-passing algorithm outperforms
commonly used state-of-the-art tools, and it nearly achieves the performance of the
aforementioned MAP decision scheme.


\end{abstract}

\IEEEpeerreviewmaketitle

\section{Introduction}

In the last decade,
rapid development of next-generation DNA sequencing technologies has enabled cheap
and fast generation of massive amounts of sequencing data \cite{Shendure08, Met05,
Bentley06}. Determining the order of nucleotides in a long target DNA molecule typically
involves use of shotgun sequencing strategy where multiple copies of the target are
fragmented into short templates. Each template is then analyzed by a sequencing instrument
which provides reads (i.e., information about nucleotide content of the templates) that are
used to assemble the desired long sequence. Shotgun sequencing strategy is
illustrated in Fig.~1.
\begin{figure}[htbp]
\centering
\includegraphics[width=0.51\textwidth]{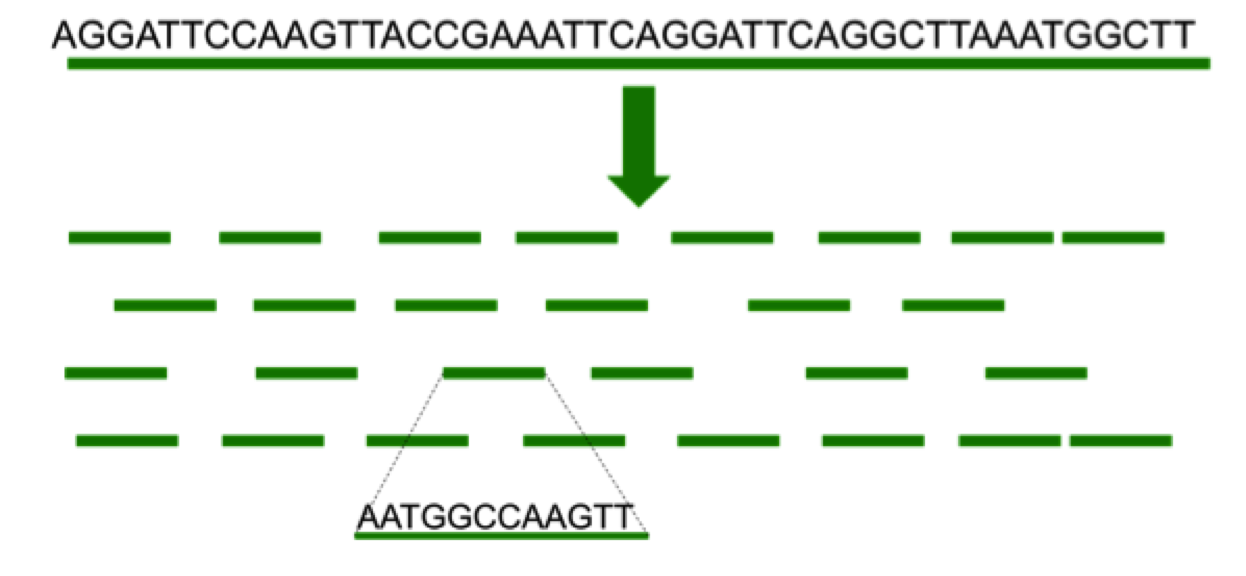}
\caption{\it Illustration of shotgun sequencing. Multiple copies of a long target DNA
molecule are fragmented into short templates. The order of bases in the
templates (reads) is determined using a sequencing instrument, and the reads are
then used to assemble the target.}
\label{fig:err}
\end{figure}

The majority of next-generation sequencing methods detect the order of nucleotides in
a template by facilitating enzymatic synthesis of a complementary strand on the
template, and acquiring a signal that indicates the type of a nucleotide successfully incorporated
into the complementary strand. Quality of the acquired signal is adversely affected by
various sources of uncertainty. Base calling algorithms attempt to infer the order of
nucleotides in short templates from the acquired noisy signals \cite{lede11}. Confidence
in the accuracy of base calls is expressed by their quality scores, which provide a measure
of the probability of base calling error. Conventional base calling algorithms rely on various
heuristics to estimate quality scores, while more recent methods employ Bayesian
inference schemes to evaluate posteriori probabilities of the bases in the reads
\cite{Song09,Vikalo12,Das13}. Due to their computational efficiency, the heuristic methods
are the preferred choice for assessing confidence of the base calls -- however, the resulting
quality scores are not necessarily an accurate reflection of the actual probability of
base calling errors \cite{Nielsen11}. In turn, inaccurate quality scores may adversely impact
reliability of the assembly process.

The short reads generated by a sequencing instrument are used to assemble the target
genome. The assembly may be performed with or without referring to a previously determined
sequence related to the target (genome, transcriptome, proteins). {\it De novo} assembly refers
to a scenario where the reconstruction is performed without a reference sequence. This is a
computationally challenging task, difficult due to the presence of perfect repeat regions in the
target sequence and short lengths of the reads \cite{Dalloul10}, \cite{Li10}. In re-sequencing
projects where the goal, for instance, may be to study genetic variations among individuals or
to discover new strains of bacteria \cite{Altmann12}, \cite{Li09b}, a reference is available and
used to order the reads. Such {\it reference-guided} assembly is still challenging due to the
errors in the reads and because the reference often contains errors and gaps \cite{Li09c},
\cite{DePristo11}. Many of the assembly challenges are ameliorated if the target sequence is
significantly oversampled and thus the information provided by short reads is highly redundant.
This redundancy is quantified by means of a sequencing coverage -- the average number of
times a base in the target sequence is present in the overlapping reads. However, the
demands for higher throughput and lower sequencing costs often limit the
coverage to medium (5-20X) or low ($\le 5$X). As an example, the ongoing {\it 1000 Genomes
Project} has opted for trading-off sequencing depth for the number of individuals being
sequenced \cite{durb1000}. In its preliminary phase, the project has focused on sequencing a
large number of individuals at a very low 3X coverage.

In the reference-guided assembly, the short reads are first mapped to a reference sequence
using an alignment algorithm (e.g., \cite{Langmead09},
\cite{Li09}). Then each position along the target is determined by combining information provided
by all the reads that cover that particular position. Due to the errors in base calls, short
length of the reads, and repetitiveness in the target, both the mapping and the sequence
assembly steps are potentially erroneous. The widely used tools to
analyze and assemble genome sequence from high-throughput sequencing data include
SAMtools \cite{Li09c} and Genome Analysis Toolkit (GATK)~\cite{DePristo11}. Note that
both of these packages rely on
the quality scores provided by the sequencing platform to infer the assembled sequence.

In this paper, we formulate the reference-guided assembly problem as the inference
of the genome sequence on a bipartite graph and solve it using a message-passing algorithm.
Unlike existing state-of-the-art methods, the proposed algorithm seeks the target sequence
without needing to know reliability of the short reads (i.e., their quality scores). Instead, it
infers reliability of a base in the assembled sequence by combining the information of all the
reads covering that particular position. The proposed algorithm
can be interpreted as the classical belief propagation under a certain prior. Binary
reformulation of the problem leads to an alternative solution in the form of another
message passing algorithm that is closely related to the so-called power iteration
method. The power iteration algorithm approximates the solution to the sequence
assembly problem by the leading singular vector of a matrix comprising read data.
The power iteration method has guaranteed convergence, and its careful examination
provides relation between the algorithm accuracy and the number of iterations.
To evaluate and benchmark performance of the proposed techniques, we find an
analytical expression for the probability of error of a genie-aided maximum a posteriori
(MAP) sequence assembly scheme which is an idealized assembler with perfect
quality score information and error-free mapping of the reads to their locations. Results
on both simulated and
experimental data obtained by sequencing {\it Escherichia Coli} and {\it Neisseria
Meningitidis} at UT Austin's Center for Genomic Sequencing and Analysis
demonstrate that our proposed message-passing algorithm performs close to the
aforementioned genie-aided MAP assembly scheme and is superior compared to
state-of-the-art methods (in particular, it outperforms the aforementioned SAMtools and GATK software
packages). Note that the developed algorithms as well as simulation
and experimental studies are focused on haploid genomes -- while modifications
that enable application to diploid/polyploid genomes are relatively straightforward,
they are beyond the scope of the current manuscript.

The paper is organized as follows. In Section \ref{model_algorithm}, we introduce
the bipartite graphical model and a message passing based sequence assembly
algorithm. In Section \ref{sec:standardbp}, we show that this message passing algorithm
can be interpreted as the classical belief propagation under a certain prior. In Section
\ref{sec:binary}, we derive an alternative message passing scheme based on a binary
reformulation of the sequence assembly problem. In Section~V, we derive an
expression for the probability of error of a genie-aided MAP
assembly scheme. In Section \ref{sec:results},
we present simulations as well as experimental results obtained by applying the proposed
methods on the {\em Escherichia Coli} and  {\it Neisseria Meningitidis} data sets. Section
\ref{sec:conclusion} concludes the paper and outlines potential future work.

Preliminary work on the basic message passing scheme discussed in Section~III was
presented in \cite{shen12,shen13}. Implementation code of the algorithm in C++ is available at https://sourceforge.net/projects/mpsequencing/.

\vspace{5mm}
\section{Graphical Model and the Message-Passing Assembly Algorithm}
\label{model_algorithm}

To facilitate processing of the short reads generated by next-generation sequencing
instruments, we introduce a bipartite graph representing the reads and bases in the target
sequence that needs to be assembled. The fundamental building blocks of a sequence
-- the nucleotides A, C, G, and T -- are numerically represented using $4$-dimensional
unit vectors containing a single non-zero component whose position indicates type of a
nucleotide. In particular, the $4$-dimensional unit vectors that we use are
$\e_A = [1 \; 0 \; 0 \; 0]^T$, $\e_C = [0 \; 1 \; 0 \; 0]^T$, $\e_G = [0 \; 0 \; 1 \; 0]^T$,
and $\e_T = [0 \; 0 \; 0 \; 1]^T$. Assume the target sequence has length $L$, and denote
the bases in the sequence by $b_{1:L}$. Then each base in the target sequence is
represented by a vector $\b_i \in \{\e_A,\e_C,\e_G,\e_T\}$. For convenience, we will assume
that all the short reads at our disposal are generated by the same sequencing instrument
and thus have identical read length $l$; note, however, that there is
no loss of generality and that our scheme can combine reads generated by sequencing
the same target on different instruments and of different read lengths. Let us denote the
set of short reads by ${\cal R} = \{r_j\}$, $j = 1, 2, \dots, n$. In general, the base calls
in these reads are erroneous due to various uncertainties in the underlying
sequencing-by-synthesis process. Average base-calling error rates of most current
next-generation sequencing systems are on the order of $10^{-2}$.

In applications where a reference sequence is available (i.e., in the so-called reference
guided sequence assembly scenario), the short reads are mapped onto the reference using
one among many recently developed short-sequence alignment algorithms
\cite{Langmead09}, \cite{Li09}. Note that the reads comprising bases with low quality scores
are often discarded by the alignment algorithms. Ideally, the remaining reads (the ones
of high fidelity) are accurately mapped to their corresponding locations on the reference
sequence. However, for some reads there may exist several candidate positions which
leads to possible mis-alignments and, consequently, might provide erroneous information
about the regions of the target sequence where the read is mis-aligned. As pointed out in
\cite{DePristo11}, the misalignment rate for reads from genome regions which contains
homozygous indels (where the chromosomes in a homologous pair have the same sequence
but contain insertion or deletion as compared to the reference genome) can be as high as $15\%$.

We represent the reference-guided assembly problem by a graph shown in Fig.~2.
\begin{figure*}
\centering
\includegraphics[width=4.65in]{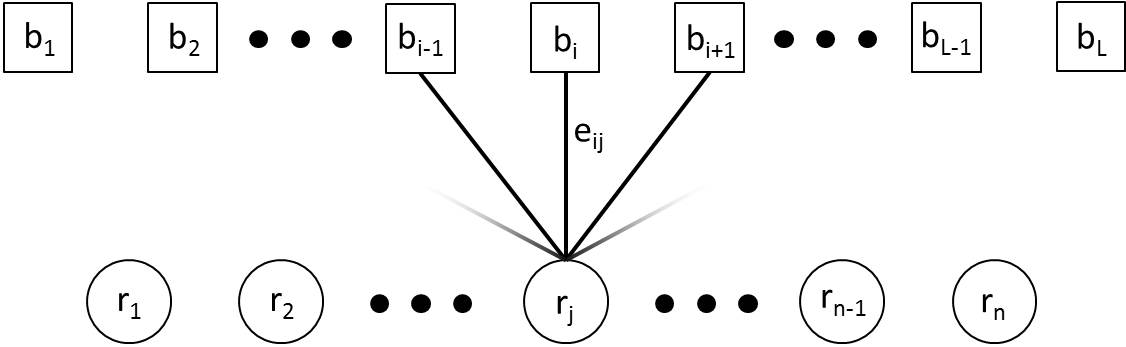}
\caption{\it Illustration of the reference-guided DNA sequence assembly problem
using short reads. Nodes $b_i$ represent bases in the target DNA sequence and $r_j$
represent reads. Each read node is connected to $l$ base nodes, where $l$ denotes
the read length.}
\label{fig:model}
\end{figure*}
The bipartite graph $G(b_{1:L}\cup r_{1:n}, E)$ illustrated in the figure has $L$ base nodes
(representing the target genome sequence) and $n$ read nodes. Since we assume that all reads are
of the same length $l$, each read node is connected to exactly $l$ base nodes. The edge
$(i,j)$ in the edge set $E$ connecting $b_i$ and $r_j$ is associated with a unit vector
$\e_{ij}$ indicating information about the type of base $b_i$ provided by the read $r_j$.
Note that the bipartite graph described here is reminiscent of the graphical representation
of the crowdsourcing problem in \cite{Shah11}. Motivated by the iterative learning scheme
proposed there, we employ a message-passing algorithm to infer the target genome
sequence using overlapping reads. Note that the previously
mentioned problem of having multiple candidate locations for mapping reads can be
incorporated in the proposed graphical representation and resolved using the algorithm
that we describe next.

The message passing algorithms rely on the exchange of messages between
neighboring nodes in the graph \cite{Kschischang01}. Our algorithm
operates on real-valued base messages $\{\x_{i\rightarrow j}\}_{(i,j)\in E}$ and
read messages $\{y_{j\rightarrow i}\}_{(i,j)\in E}$. A base message $\x_{i\rightarrow j}$
is a $4 \times 1$ vector representing the likelihood of the base $b_i$ being A, C, G, or
T, while a read message $y_{j\rightarrow i}$ represents the reliability of read $j$.
Read messages are initialized from a random distribution, and the
message update rules at iteration $k$ are given by
\begin{eqnarray}
\label{xiter}
\x^{(k)}_{i\rightarrow j}&\leftarrow &\sum_{j'\in \partial i\backslash j} (2\e_{ij'}-\mathds{1}) y_{j' \rightarrow i}^{(k-1)}, \\
\label{yiter}
y_{j\rightarrow i}^{(k)}&\leftarrow &\frac{1}{l-1}\sum_{i'\in \partial j\backslash i} \e_{i'j}^T \x_{i'
\rightarrow j}^{(k)},
\end{eqnarray}
where $\partial i$ and $\partial j$
denote collection of the neighboring nodes of nodes $i$ and $j$, respectively, and
$\mathds{1}$ is a $4\times 1$ vector containing all $1$'s. Note that
$2\e_{ij'}-\mathds{1}$ has element $1$ in the position corresponding to the nucleotide
base $b_{ij'}$ represented by $\e_{ij'}$, and $-1$'s elsewhere. Hence a read with positive
reliability value $y_{j'\rightarrow i}$ will increase the likelihood of $b_{ij'}$ and decrease
the likelihood of other bases. Finally, the likelihood of a base being A, C, G, or T is calculated
as the sum of the information provided by the reads weighted by each read's
reliability. The symbol with the highest likelihood is chosen as the estimate of the base
in the corresponding position. The estimate rule for the $i^{th}$ base is
\begin{equation}
\hat{b}_i = \arg\max_{t\in \{A,C,G,T\}} \x_i^{\{t\}},
\end{equation}
where the decision vector $\x_{i} = \sum_{j\in \partial i} (2\e_{ij}-\mathds{1})
y_{j \rightarrow i}^{(k_m)}$. Here, $k_m$ denotes the number of iterations performed
and $\x_i^{\{t\}}$ denotes the likelihood corresponding to symbol $t \in
\{A,C,G,T\}$ in the vector $\x_i = \left[\x_i^{\{A\}} \; \x_i^{\{C\}} \; \x_i^{\{G\}} \;
\x_i^{\{T\}}\right]^T$. The procedure is formalized as Algorithm~1.

\begin{algorithm}
\caption{Message passing for sequence assembly}
\label{mpsq}
\begin{algorithmic}
\STATE Input: $E$, $\{\e_{ij}\}_{(i,j)\in E}$
\STATE 1 Initialize read messages:
\FORALL{$(i,j)\in E$}
\STATE Initialize $y_{j\rightarrow i}^{(0)}$;
\ENDFOR

\STATE 2 Iterations:
\FOR{$k =1 \to k_m$}
 \FORALL{$(i,j)\in E$}
 \STATE Update base message:
 \STATE $\x_{i\rightarrow j}^{(k)}\leftarrow \sum_{j'\in \partial i\backslash j} (2\e_{ij'}-\mathds{1}) y_{j' \rightarrow i}^{(k-1)}$;
 \STATE Normalize $\x_{i\rightarrow j}^{(k)}$:
 \STATE $\x_{i\rightarrow j}^{(k)}\leftarrow\frac{\x_{i\rightarrow j}^{(k)}}{||\x_{i\rightarrow j}^{(k)}||_2}.$
 \ENDFOR

 \FORALL{$(i,j)\in E$}  \STATE Update read message
 \STATE $y_{j\rightarrow i}^{(k)}\leftarrow \frac{1}{l-1}\sum_{i'\in \partial j\backslash i} \e_{i'j}^T \x_{i' \rightarrow j}^{(k)}$; \ENDFOR
\ENDFOR

\STATE 3 Estimation:
\FOR{$i =1 \to L$}
\STATE Calculate decision vector $\x_{i} = \sum_{j\in \partial i} (2\e_{ij}-\mathds{1}) y_{j \rightarrow i}^{(k_m)}$;
\STATE Estimate the bases
\STATE {\bf $\hat{b_i} = \arg\max_{t\in \{A,C,G,T\}} \x_i^{\{t\}}$;}
\ENDFOR
\end{algorithmic}
\end{algorithm}

Note that Algorithm~1 needs to be appropriately initialized. In our experimental studies presented
in Section~VI, we initialize $y_{j\rightarrow i}^{(0)}$ by drawing from both Gaussian distribution
$\mathcal{N}(1,1)$ and uniform distribution $U[0,1]$. For the data sets under consideration, it
turns out that different initializations lead to identical solutions. The algorithm is terminated when
the reliability increment between subsequent iterations is small, i.e.,
$\sum |y_{j \rightarrow i}^{(k)}-y_{j \rightarrow i}^{(k-1)}|<\epsilon$. As pointed out earlier, the
algorithm does not require exact knowledge of quality scores, and iteratively infers reliability of
individual reads.

Since the reads originating from a single sequencing instrument have identical lengths, the degree
of the read nodes in the graph is uniform. On the other hand, degree of a base node is the number
of reads that cover the corresponding base, usually referred to as the {\em sequencing coverage}.
Typically, coverage varies from one position to another and, consequently, degree of the base
nodes varies. Note that fragmentation of multiple copies of the target sequence -- a fundamental
step in shotgun sequencing procedure -- can be viewed as a uniform sampling from the original
DNA strand. The resulting coverage is a
random variable that can be described well by a Poisson distribution \cite{kao11, Waterman2}.

Let $\bar{c}$ denote the average sequencing coverage. The computational complexity of the
base message updating step (\ref{xiter}), which needs to be performed in each iteration of
Algorithm~1, is $\mathcal{O}(nl\bar{c})$ on average, while the complexity of the read message
updating step (\ref{yiter}) is $\mathcal{O}(nl^2)$. Since $L\bar{c}=nl$, the complexity of the algorithm
is $\mathcal{O}(k_mnl(l+\bar{c}))=\mathcal{O} (k_mL\bar{c}(l+\bar{c}))$, where $k_m$ denotes
the number of iterations (i.e., the number of message updates). On the other hand,
simple plurality voting scheme has complexity $\mathcal{O}(L\bar{c}\log\bar{c})$. Our
experimental studies show that $k_m \le 30$ is sufficient for the convergence of the
algorithm. We tested the algorithm on a broad range of parameters (in particular, for read
lengths $l \le100$, coverage $\bar{c} \le 60$), and found that the runtimes are comparable
to those of the state-of-the-art techniques (SAMtools and GATK) -- a specific comparison of
runtimes is reported in Section VI.

\vspace{5mm}
\section{Relation to standard belief propagation}
\label{sec:standardbp}

As an alternative to the intuitively pleasing but basically heuristic message passing
scheme proposed in Section~\ref{model_algorithm}, we can also derive a standard belief
propagation algorithm for the reference-guided sequence assembly. To this end, we seek
the sequence $\hat{b}_{1:L}$ that maximizes the joint probability $P(\hat{b}_{1:L},p_{1:n})$,
where $p_{1:n}$ denotes confidences of the aligned read data and $p_j\in [0,1]$. This
maximization can be formalized as
\begin{equation}
\label{optimization}
\max_{\hat{b}_{1:L},p_{1:n}} \prod_{j=1}^n \mathcal{D}(p_j) \prod_{(i,j)\in E} \left\{ p_j
\delta(\hat{b}_i=\e_{ij}) +\bar{p}_j\delta(\hat{b}_i\neq\e_{ij}) \right\},
\end{equation}
where $\mathcal{D}(p_i)$ denotes the prior distribution on $p_i$ and $\bar{p}_j=1-p_j$. $\delta(\cdot)$ denotes
an indicator function taking value $1$ if its
argument is true and is $0$ otherwise.
The joint optimization is computationally challenging and thus often practically not
feasible. As an alternative, belief propagation provides an approximate solution to
(\ref{optimization}) by computing the marginal distributions of the optimization variables
and selecting their most likely values according to the computed distributions. A thorough
review of theoretical and practical aspects of the belief propagation method can be found
in \cite{YFW05}. For the graphical model proposed in Section~\ref{model_algorithm}, we
define two messages to facilitate belief propagation: $\tilde{x}_{i\rightarrow j}$ and
$\tilde{y}_{j\rightarrow i}$. The former is the belief on $\hat{b}_i$ and essentially represents
a distribution over the four possible nucleotide bases $\{\e_A,\e_C,\e_G,\e_T\}$. The
latter is a probability of $p_j$ on $[0,1]$. In the $k^{th}$ iteration of the belief
propagation algorithm, message update rules are given by (see, e.g., \cite{YFW05} and
the references therein)
\begin{eqnarray}
\label{bpy}
\tilde{y}_{j\rightarrow i}^{(k)}(p_j)&\propto &\mathcal{D}(p_j) \prod_{i'\in \partial j\setminus i} \sum_{m=A,C,G,T}\Big\{ p_j \delta(\e_{i'j}=\e_m)  \nonumber\\
&& +\bar{p}_j\delta(\e_{i'j}\neq \e_m) \Big\} \tilde{x}_{i'\rightarrow j}^{(k)}(\e_m), \\
\label{bpx}
\tilde{x}_{i\rightarrow j}^{(k+1)}(\hat{b}_i)&\propto &\prod_{j'\in \partial i\setminus j}\int \Big(\tilde{y}_{j'\rightarrow i}^{(k)}(p_{j'}) (p_{j'} \delta(\hat{b}_i=\e_{ij'}) \nonumber\\
&&+\bar{p}_{j'}\delta(\hat{b}_i\neq\e_{ij'})\Big)d p_{j'}.
\end{eqnarray}
After the completion of the iterative procedure, the bases $b_i$ in the target genome are
estimated by first computing the beliefs $\tilde{x}_i(\hat{b}_i) \propto$
\begin{equation}
\prod_{j'\in \partial i}\int \Big(\tilde{y}_{j'\rightarrow i}^{(k)}(p_{j'}) (p_{j'} \delta(\hat{b}_i=\e_{ij'}) +\bar{p}_{j'}\delta(\hat{b}_i\neq\e_{ij'}))\Big)d p_{j'},
\end{equation}
where $\hat{b}_i\in\{\e_A,\e_C,\e_G,\e_T\}$, and then choosing the base with the highest
$\tilde{x}_i$ value. Note that, by exploiting the symmetry of the expression (\ref{optimization}),
we can write
\begin{eqnarray}
\tilde{x}_{i\rightarrow j}^{(k+1)}(\hat{b}_i\neq \e_m) &\propto & \prod_{j'\in \partial i\setminus j}\int \Big(\tilde{y}_{j'\rightarrow i}^{(k)}(p_{j'}) (p_{j'} \delta(\e_{ij'}\neq \e_m)\nonumber \\
 & &+\bar{p}_{j'}\delta(\e_{ij'}=\e_m)\Big)d p_{j'}. \nonumber
\end{eqnarray}
For the brevity of notation, we denote
$\tilde{x}_{i\rightarrow j}^{(k)}(\e_m)=\tilde{x}_{i\rightarrow j}^{(k)}(\hat{b}_i= \e_m)$.
Assuming that the prior distribution on $p_j$, $\mathcal{D}(p_j)$, is Beta(0,0) (which is
essentially as same as the Bernoulli(1/2) distribution), the read confidence is a binary variable,
\[
p_j = \left\{
\begin{array}{cc}
0, \mbox{ w.p. } 1/2 ,\\
1, \mbox{ w.p. } 1/2.
\end{array}
\right.
\]
Define a log-likelihood ratio
\begin{equation}
\label{ylikelihood}
y_{j\rightarrow i}^{k}=\log\Big(\frac{\tilde{y}_{j\rightarrow i}^{(k)}(1)}{\tilde{y}_{j\rightarrow i}^{(k)}(0)}\Big).
\end{equation}
After substituting (\ref{bpy}) in (\ref{ylikelihood}), we obtain
\begin{eqnarray}
y_{j\rightarrow i}^{(k)} &=& \sum_{i'\in \partial j\setminus i}\log \frac{\tilde{x}_{i'\rightarrow j}^{(k)}(\e_{i'j})}{\sum_{\e_m\neq \e_{i'j}}\tilde{x}_{i'\rightarrow j}^{(k)}(\e_m)} \nonumber \\
&=& \sum_{i'\in \partial j\setminus i} \log \frac{\tilde{x}_{i'\rightarrow j}^{(k)}(\e_{i'j})}{\tilde{x}_{i'\rightarrow j}^{(k)}(\hat{b}_{i'}\neq \e_{i'j})}.
\end{eqnarray}
Define a $4\times 1$ vector message $\x_{i\rightarrow j}^{(k)}$ as
\[
\x_{i\rightarrow j}^{(k)} = \left[
\x_{i\rightarrow j}^{(k)}(1) \;\;\; \x_{i\rightarrow j}^{(k)}(2) \;\;\;
\x_{i\rightarrow j}^{(k)}(3) \;\;\; \x_{i\rightarrow j}^{(k)}(4) \right]^T,
\]
where
\[
\begin{array}{lr}
\x_{i\rightarrow j}^{(k)}(1) = \log \frac{\tilde{x}_{i\rightarrow j}^{(k)}(\e_{A})}{\tilde{x}_{i\rightarrow j}^{(k)}(\hat{b}_{i}\neq \e_{A})}, \;
\x_{i\rightarrow j}^{(k)}(2) = \log \frac{\tilde{x}_{i\rightarrow j}^{(k)}(\e_{C})}{\tilde{x}_{i\rightarrow j}^{(k)}(\hat{b}_{i}\neq \e_{C})}, \\
\x_{i\rightarrow j}^{(k)}(3) = \log \frac{\tilde{x}_{i\rightarrow j}^{(k)}(\e_{G})}{\tilde{x}_{i'\rightarrow j}^{(k)}(\hat{b}_{i'}\neq \e_{G})}, \;
\x_{i\rightarrow j}^{(k)}(4) = \log \frac{\tilde{x}_{i\rightarrow j}^{(k)}(\e_{T})}{\tilde{x}_{i'\rightarrow j}^{(k)}(\hat{b}_{i'}\neq \e_{T})}.
\end{array}
\]
It is straightforward to write
\begin{equation}
\label{ybp}
y_{j\rightarrow i}^{(k)}=\sum_{i'\in \partial j\setminus i} \e_{i'j}\x_{i'\rightarrow j}^{(k)}.
\end{equation}
A closer examination of the first element of $\x_{i\rightarrow j}^{(k)}$,
$\x_{i\rightarrow j}^{(k)}(1)$, leads to simplification shown in (\ref{xderive}), where we
implicitly used the assumption that $p_j$ is binary.
\begin{table*}[t]
\begin{eqnarray}
\x_{i\rightarrow j}^{(k)}(1) & = &
\log\frac{x_{i\rightarrow j}^{(k)}(\e_A)}{x_{i\rightarrow j}^{(k)}(\hat{b}_i\neq\e_A)}
= \sum_{j'\in\partial i\setminus j}\log\frac{\int\left(\tilde{y}_{j'\rightarrow i}^{(k-1)}(p_{j'}) (p_{j'} \delta(\e_{ij'}=\e_A) +\bar{p}_{j'}\delta(\e_{ij'}\neq \e_A)\right)d p_{j'}}
{\int\left(\tilde{y}_{j'\rightarrow i}^{(k-1)}(p_{j'}) (p_{j'} \delta(\e_{ij'}\neq \e_A) +\bar{p}_{j'}\delta(\e_{ij'}=\e_A)\right)d p_{j'}} \nonumber \\
& = & \left\{
 \begin{array}{ll}
\log\frac{\tilde{y}_{j'\rightarrow i}^{(k-1)}(1)}{\tilde{y}_{j'\rightarrow i}^{(k)}(0)}=y_{j'\rightarrow i}^{(k-1)}& \mathrm{if}\; \e_{ij'}=\e_A \\
-\log\frac{\tilde{y}_{j'\rightarrow i}^{(k-1)}(1)}{\tilde{y}_{j'\rightarrow i}^{(k)}(0)}=-y_{j'\rightarrow i}^{(k-1)}& \mathrm{if}\;\e_{ij'}\neq \e_A\end{array}\right.
\label{xderive}
\end{eqnarray}
\hrule
\end{table*}
We can obtain similar expressions to (\ref{xderive}) for other components of
$\x_{i\rightarrow j}^{(k)}$. As a result, the updating rule for $\x_{i\rightarrow j}^{(k)}$
simplifies,
\begin{equation}
\label{xbp}
\x_{i\rightarrow j}^{(k)} = \sum_{j'\in \partial i\setminus j} (2\e_{ij'}-\mathds{1})
y_{j' \rightarrow i}^{(k-1)},
\end{equation}
where the vector $2\e_{ij'}-\mathds{1}$ has element $1$ in the position corresponding to the
nucleotide base $b_{ij'}$ represented by $\e_{ij'}$, and $-1$'s elsewhere. Therefore, the belief
propagation update rule (\ref{xbp}) is identical to the update rule (\ref{xiter}) of our message
passing algorithm presented in Section~II. Moreover, update rule (\ref{ybp}) is identical (up to
the scaling factor) to the message update rule (\ref{yiter}). Therefore, message passing scheme
proposed in Section~II can be interpreted as the belief propagation under a specific prior
on the confidence of the aligned data $p_j$ -- in particular, $p_j$ should come from a
Beta(0,0) distribution, i.e., be treated as a binary variable.

\vspace{5mm}
\section{Binary representation, message passing, and power iteration algorithm}
\label{sec:binary}

So far, we discussed reference-guided assembly schemes that rely on a representation of the
nucleotide basis with $4$-dimensional vectors $\{\e_A,\e_C,\e_G,\e_T\}$. As an
alternative, in this section we rely on a binary representation of nucleotides to formulate
a message passing scheme and discuss the provably convergent power iteration
algorithm for finding the target genome sequence. The power iteration scheme finds the desired
sequence by computing the leading singular vectors of an appropriately defined data matrix.

The four-letter alphabet $\{A,C,G,T\}$ in DNA sequencing data can be represented using binary symbols, e.g.,
$\{+1,-1\}$. In particular, we encode the nucleotide basis as $A=\{-1,-1\}$, $C=\{-1,+1\}$,
$G=\{+1,-1\}$, and $T=\{+1,+1\}$, and represent reads as binary sequences comprising
$\{\pm 1\}$. Similar to how we built a model utilizing $4$-dimensional vectors
$\{\e_A,\e_C,\e_G,\e_T\}$ in Section~II, we define a bipartite graph where each base $b_i$
is represented by two binary nodes $\tilde{b}_{2i-1}$ and $\tilde{b}_{2i}$. Using the
output of an alignment algorithm, each read node of the bipartite graph is connected to $2l$
binary base nodes in the node set $\tilde{b}_{1:2L}$, where $l$ denotes read length and
$L$ is the length of the target sequence. For convenience, let us denote the resulting
graph by $G(\tilde{b}_{1:2L}\cup r_{1:n},\tilde{E})$. The edge $(k,j)$ in $\tilde{E}$ connecting
$\tilde{b}_k$ and $r_j$ is assigned a variable $e_{kj}\in\{\pm 1\}$, the binary representation
of $\tilde{b}_k$ provided by read $r_j$. Given such a graphical representation, we can apply a
binary message passing algorithm as in \cite{Shah11}. In particular, the read and base
messages are scalars and the update equations are given by
\begin{eqnarray}
\label{xiter2}
x^{(k)}_{i\rightarrow j}&\leftarrow &\sum_{j'\in \partial i\backslash j} e_{ij'} y_{j' \rightarrow i}^{(k-1)}, \\
\label{yiter2}
y_{j\rightarrow i}^{(k)}&\leftarrow &\sum_{i'\in \partial j\backslash i} e_{i'j}^T x_{i'
\rightarrow j}^{(k)}.
\end{eqnarray}
After the iterative procedure reaches a stopping criterion, the binary string representing
unknown target DNA sequence is obtained as the weighted average
\begin{equation}
\tilde{b}_i = \mathrm{sign}(\sum_{j'\in \partial i\backslash j} e_{ij'} y_{j' \rightarrow i}).
\end{equation}
The above algorithm is known to converge to the optimal solution when the bi-partite graph
is regular \cite{Shah112}. In our application, however, the graph is not regular since the
sequencing coverage varies. Nevertheless, we find that the binary message-passing algorithm performs
very well in both simulations and on experimental data, as we demonstrate in Section~VI.
The binary message passing algorithm is also closely related to the so-called power iteration
scheme for computing the leading singular vector of an appropriately defined data matrix.
We next examine the power iteration algorithm and argue its convergence.

With the adopted binary encoding of nucleotides, we can represent sequencing reads by
a sparse $n \times 2L$ matrix $D$. The $2L$ columns of $D$ correspond to the $L$ positions
in the target sequence whereas the $j^{th}$ row of $D$ comprises binary data representing
read $r_j$. In each row, only $2l$ entries are non-zero (representing an $l$-long read) while
the remaining ones are filled with zeros. Therefore, matrix $D$ has entries $D_{ij}\in\{0,+1,-1\}$.
Since the percentage of nonzero entries of $D$ is $\frac{2l}{L}$ and $L\gg l$, $D$ is a sparse matrix.
It is easy to show (see, e.g., \cite{Shah112}) that if each row of $D$ has the same number of
nonzero entries, and the same holds for each column, the left singular vector corresponding
to the largest singular value of $D$ is a reliable estimate of the target genome sequence when
the measurement noise (i.e., read error rate) is low.
Here is an illustration. Let $s$ denote the $2L\times 1$ binary
vector with alphabet $\{-1,+1\}$ representing the true sequence of length $L$, and let the number of nonzero entries in
each columns of $D$ be $c$. Consider the case where the reads are error-free and $s$ is a $2L\times 1$ all one vector $\mathds{1}_{2L}$.
Since $DD^T \mathds{1}_{2L}=2Lc \mathds{1}_{2L}$,
then $s$ is an eigenvector of $DD^T$. Here $D$ is a non-negative matrix with
entries $0$s and $1$s and thus, by Perron-Frobenius theorem, $\mathds{1}_{2L}$ is a left
singular vector corresponding to $D$'s largest singular value. In the general case where $s$
consists of both $1$ and $-1$, we can represent $s=S\mathds{1}_{2L}$ where $S$ is a
$2L\times 2L$ diagonal matrix with $\mathrm{diag}(S)=s$. In this case, it is straightforward
to generalize the above analysis and show that $s$ remains to be proportional to the leading
singular vector of the matrix $D$.

Performing singular value decomposition is roughly cubic in the dimension of $D$ and,
for our problem dimensions, clearly infeasible. Fortunately, we only need to find $\u$,
the leading singular vector of $D$, and then estimate the target sequence $s$ as
$\mathrm{sign}(\u)$. This can be done in a computationally efficient way using the power
iteration technique due to sparsity of $D$. In particular, the power iteration procedure
entails computing
\begin{equation}
\label{poweriter}
\x^{(k)}=D\y^{(k-1)}, \;\;\; \y^{(k)}=D^T \x^{(k)}.
\end{equation}
To demonstrate convergence of the power iteration scheme (\ref{poweriter}), let us denote
the singular values of $D$ as $\sigma_i(D)$, where
$\sigma_1(D)\geq\sigma_2(D)\geq...\geq 0$. With a random initialization $\y^{(0)}$, power
iterations will converge to the singular vector $\u$ if the inequality $\sigma_1(D)>\sigma_2(D)$
holds strictly. The speed of the convergence of power iterations depends on the ratio
$\sigma_2(D)/\sigma_1(D)$. This can be easily shown by an analysis of the consecutive
projections of the iteratively updated vectors $\x^{(k)}$ onto the singular vector $\u$. In
particular, the projection of $\x^{(k)}$ onto $\u$ is $(\u^T\x^{(k)})\u$. A closer look into the
singular value decomposition shows that $\u^T\x^{(k)}\u=(\sigma_1(D))^2\u^T\x^{(k-1)}\u$
and $(\x^{(k)}-\u^T\x^{(k)}\u)\leq (\sigma_2(D))^2 (\x^{(k-1)}-\u^T\x^{(k-1)}\u)$. Therefore,
\begin{eqnarray*}
\frac{||\x^{(k)}-\u^T\x^{(k)}\u||}{||\u^T\x^{(k)}\u||} &\leq& \left(\frac{\sigma_2(D)}{\sigma_1(D)}
\right)^2 \frac{||\x^{(k-1)}-\u^T\x^{(k-1)}\u||}{||\u^T\x^{(k-1)}\u||}\nonumber \\
&\leq & \left(\frac{\sigma_2(D)}{\sigma_1(D)} \right)^{2k} \frac{||\x^{(0)}-\u^T\x^{(0)}\u||}{||
\u^T\x^{(0)}\u||}.
\end{eqnarray*}
Clearly, power iterations will converge with any initialization if $\sigma_1(D)  >\sigma_2(D)$,
and the speed of convergence depends on the ratio of $\sigma_1(D)$ and $\sigma_2(D)$ --
the larger the ratio, the faster the convergence. On the other hand, from (\ref{poweriter}) it
directly follows that the update equations for the entries of $\x^{(k)}$ and $\y^{(k)}$ can be
written as
\begin{equation}
x_i^{(k)} = \sum_{j\in\partial i}D_{ij} y_j^{(k-1)}, \;\;\;
y_j^{(k)} = \sum_{i\in\partial j}D_{ij} x_i^{(k)}.
\label{PIcomp}
\end{equation}
Note that the power iterations (\ref{PIcomp}) differ from the message update rules (\ref{xiter2})
and (\ref{yiter2}) in only one term. As our results in Section~VI show, accuracy of message
passing and power iterations is essentially identical, while the former converges in significantly
fewer iterations than the latter. Moreover, both message-passing schemes -- the one
based on the representation of basis via $4$-dimensional vectors
$\{\e_A,\e_C,\e_G,\e_T\}$ as well as the one relying on the binary representation of nucleotides
-- converge after approximately the same number of iterations.

\vspace{5mm}
\section{Benchmarking performance of the proposed assembly schemes}
\label{sec:benchmark}

To assess and benchmark performance of the proposed iterative learning schemes,
in this section we analyze the probability of error of a genie-aided maximum a posteriori
(MAP) estimator of the bases in the target genome sequence. In this problem, "genie-aided" is referring to an
idealized scenario where short reads are mapped to the reference genome with no
errors, i.e., there are no misplacements of the reads along the reference sequence
and the MAP estimator knows exact probabilities of
mis-calling the bases in the short reads (i.e., has exact quality score information).
Recall that neither our message-passing schemes nor the power iteration algorithm
make such practically unrealistic assumptions and, in fact, do not require prior
knowledge of quality scores.

\subsection{Genie-aided MAP estimator}

Let $b_k$ denote the $k^{th}$ base in the target sequence, and let $y^{(i)}_k$
denote the signal generated by the sequencing platform as it examines $b_k$,
$i = 1,2,\ldots, c_k$, where $c_k$ stands for the total number of reads covering
$b_k$. Assume that the probability of erroneously calling $b_k$ in the $i^{th}$
read is $p^{(i)}_k$. Given the base calls of the reads covering $b_k$, $y^{(i)}_k$,
the MAP estimate $\hat{b}_k$ is readily found as
\begin{eqnarray}
\hat{b}_k &=& \arg\max_{x} \prod_{i=1}^{c_k} P(y^{(i)}_k,b_k=x)\nonumber\\
&=&\arg\max_{x}\prod_{i=1}^{c_k}P(y^{(i)}_k|b_k=x)P(b_k=x)\nonumber\\
&=&\arg\max_{x}\prod_{i=1}^{c_k}(1-p^{(i)}_k)^{\delta(y^{(i)}_k=x)}
(p^{(i)}_k)^{(1-\delta(y^{(i)}_k=x))} P(b_k = x) \nonumber \\
&=&\arg\max_{x}\sum_{i=1}^{c_k}\Big[\delta(y^{(i)}_k=x)\log(1-p^{(i)}_k) \nonumber \\
&+&(1-\delta(y^{(i)}_k=x))\log(p^{(i)}_k)\Big]+\log(P(b_k=x))\nonumber\\
&=&\arg\max_{x}\sum_{i=1}^{c_k}\delta(y^{(i)}_k=x) w^{(i)}_k +
\log(p^{(i)}_k) \nonumber \\ &+& \log(P(b_k=x)), \nonumber
\end{eqnarray}
where we introduced
$w^{(i)}_k=\log\Big(\frac{1-p^{(i)}_k}{p^{(i)}_k}\Big)$. Therefore, the MAP estimate
$\hat{b}_k$ formed by combining the information provided by $c_k$ reads
covering $b_k$ is given by
\begin{equation}
\hat{b}_k=\arg\max_{x}\sum_{i=1}^{c_k} w^{(i)}_k\delta(y^{(i)}_k=x)+\log(P(b_k=x)).
\label{mapalgo}
\end{equation}
In the absence of prior information $P(b_k=x)$, the MAP estimation of $b_k$ in
(\ref{mapalgo}) is identical to the so-called weighted plurality voting  \cite{lin02}.
Note that if $p^{(i)}_k = p$ for all $i$ and $k$, (\ref{mapalgo}) becomes the
well-known plurality voting scheme.

Intuitively, we expect that the performance of the MAP decision scheme improves
as we increase the coverage $c_k$. Note that the above expressions are predicated
on the assumption of error-free read mapping.

\subsection{Performance of the genie-aided MAP estimator}
For notational convenience, let us write the expression for the estimate in (\ref{mapalgo})
as
\begin{equation}
\hat{b}_k=\arg\max_{x} W_k(x),
\end{equation}
where $W_k(x)=\displaystyle\sum_{i:y^{(i)}_k=x} w^{(i)}_k+\log(P(b_k=x))$.
\footnote{Without a loss of generality, we will assume that ties where two
different bases $x$ and $y$ lead to identical $W_k(x)=W_k(y)$ do not happen.
The extension to this case is trivial but requires more cumbersome notation.}
The probability of error is defined as $P(\hat{b}_k \neq b_k)=1-P(\hat{b}_k = b_k)$.

To characterize the probability of error of the MAP decision scheme, we rely on
the so-called universal generating functions often used in reliability analysis of
multi-state systems \cite{Levitin}. Consider $n$ independent discrete random variables
$X_1, \ldots, X_n$ with probability mass functions (pmf)
represented by vectors $(\bm{x}_i, \bm{p}_i)$ (e.g., $P(X_i=x_{ij})=p_{ij}$).  In order to
evaluate the pmf of an arbitrary function $f(X_1,\ldots,X_N)$, one has to find the
vector $\bm{y}$ of all the possible values of $f(\cdot)$ and the vector $\bm{q}$ of
the corresponding probabilities. The total number
of possible combinations $(X_1,\ldots,X_N)$ is $K=\prod_{i=1}^{n}(k_i+1)$,
where $(k_i+1)$ is the number of different realizations of $X_i$. Since the
variables are independent, the probability of each unique combination is
equal to the product of the probabilities of the realizations of arguments composing
this combination. The probability of the $j^{th}$ combination of the realizations of
the variables is $q_j=\prod_{i=1}^{n}p_{ij}$ and the
corresponding value of the function is $f_j=f(x_{1j_1},\ldots,x_{nj_n})$.
If different combinations produce the same value of the function, then the
probability that $f(\cdot)$ takes that value is equal to the sum of probabilities of the
combinations resulting in it. As an illustration, let $A_h$ denote the set of
combinations resulting in the particular function value $f_h$. If the total number of different
values that the function of random variables $f(X_1,\ldots,X_n)$ may assume is $H$,
then its probability mass function is completely specified with a pair of vectors
$(\bm{y},\bm{q})$ defined as
\begin{eqnarray*}
\bm{y} &=& (f_h:1\leq h \leq H), \\
\bm{q} &=& \Big(\sum_{(x_{1j_1},\ldots,x_{nj_n}) \in A_h}\prod_{i=1}^{n}p_{ij_i}:1\leq h \leq H \Big).
\end{eqnarray*}
A compact representation of the probability mass function of a random variable $X_i$,
$(x_{i0},x_{i1},\ldots,x_{ik_i},p_{i0},p_{i1},\ldots,p_{ik_i})$, is given by a z-transform that
takes the polynomial form
\begin{equation}
u_i(z)=\sum_{j=0}^{k_i}p_{ij}z^{x_{ij}}.
\label{ztransform}
\end{equation}
Such a representation is convenient since, to find the probability that $X_i \in \Phi$, one can
use operator $\delta$ defined as
\begin{equation}
Pr(X_i \in \Phi)=\delta(u_i(z),\Phi)=\sum_{j:x_{ij} \in \Phi}p_{ij}.
\end{equation}
Moreover, the z-transform representation enables straightforward calculation of the probability
mass function of an arbitrary function $f$ of $n$ independent random variables. This can be
facilitated via the composition operator ${\otimes}_f$ applied to z-transform representations of
the probability mass functions of the variables,
\begin{equation}
{\otimes}_f\Big(\sum_{j_i=0}^{k_i}p_{ij_i}z^{x_{ij_i}}\Big)
=\sum_{j_1=0}^{k_1}\ldots\sum_{j_1=0}^{k_n}\Big(\prod_{i=0}^{n}p_{ij_i}z^{f(x_{1j_1},
\ldots,x_{nj_n})}\Big).
\end{equation}
The technique for finding probability mass functions that relies on the z-transform and
composition operators ${\otimes}_f$ is referred to as the universal z-transform or the
universal (moment) generating function (UGF) technique. In the context of this technique,
the z-transform of a random variable for which the operator ${\otimes}_f$ is defined is
often referred to as its U-function. For additional background on this subject, we refer
an interested reader to \cite{Levitin}. Here, we rely on this technique to characterize the
probability of error of the genie-aided sequence assembly scheme.

Consider the U-function (similar to (\ref{ztransform})) defined for each read position
\begin{eqnarray}
U_{i}(z)&=&\sum_{m=1}^{4}s^{i}_{m}z^{v^{i}_{m}}\nonumber\\
&=&r^{(i)}_{x_1}z^{[w_i(x_1)\hspace{0.5mm} 0\hspace{0.5mm} 0\hspace{0.5mm} 0]}+r^{(i)}_{x_2}z^{[0\hspace{0.5mm} w_i(x_2)\hspace{0.5mm} 0 \hspace{0.5mm}0]} \nonumber \\ &+& r^{(i)}_{x_3}z^{[0\hspace{0.5mm} 0 \hspace{0.5mm}w_i(x_3)\hspace{0.5mm} 0]}+r^{(i)}_{x_3}z^{[0\hspace{0.5mm} 0\hspace{0.5mm} 0 \hspace{0.5mm}w_i(x_4)]},
\label{HP1}
\end{eqnarray}
where $r^{i}_{x_j}$ denotes the probability that the symbol from read $i$ is $x_j$, and
$w_i(x_j)$ is the weight associated with the information provided by read $i$ (essentially
given by the quality scores, which the genie-aided scheme assumes to be perfectly known).
To obtain a U-function of the decision for two positions having respective U-functions $U_{1}(z)$
and $U_{2}(z)$, the following composition operator can be used,
\begin{eqnarray}
U_{1,2}(z)&=&\Omega\Big(U_1(z), U_2(z)\Big)\nonumber\\
&=&\Omega\Big(\sum_{m=1}^{4}s^{1}_{m}z^{v^{1}_{m}}, \sum_{m=1}^{4}s^{2}_{m}z^{v^{2}_{m}}\Big)\nonumber\\
&=&\sum_{m=1}^{4}\sum_{n=1}^{4}s^{1}_{m}s^{2}_{n}z^{v^{1}_{m}+v^{2}_{m}}
\nonumber \\
&=& \sum_{m}^{}s^{\{1,2\}}_{m}z^{v^{\{1,2\}}_{m}}.
\label{twoterms1}
\end{eqnarray}
Note that some combinations of $v^{1}_{m}$ and $v^{2}_{m}$ may lead to the same
$v^{1}_{m}+v^{2}_{m}$ and hence there may be multiple terms in (\ref{twoterms1})
that involve $z^{v^{1}_{m}+v^{2}_{m}}$. If so, in the last step in (\ref{twoterms1}),
such terms are summed up to obtain $s^{\{1,2\}}_{m}$ which is referred to as the
read output distribution of reads 1 and 2. The support set of $s^{\{1,2\}}_{m}$ is
at most $4^2 = 16$, but may be smaller due to aforementioned grouping of the
terms that involve identical vectors.

The above procedure leads to a representation of the probability of error of arriving
at the decision for a particular sequence position by combining information provided
by two reads. Given an
arbitrary subset of reads $\lambda$ (e.g., so far we discussed $\lambda=\{1,2\}$),
it is straightforward to obtain the U-function for an extended subset $\lambda \cup {j}$ with
an arbitrary $j \notin \lambda$ as
\begin{eqnarray}
U_{\lambda\cup j}(z)&=&\Omega\Big(U_{\lambda}(z), U_j(z)\Big)
=\sum_{m}^{}s^{\lambda\cup j}_{m}z^{v^{\lambda\cup j}_{m}}.
\label{generalterm}
\end{eqnarray}
We can further simplify and arrive at more explicit expressions in the following way.
Consider the U-function of an arbitrary weighted voting classifier (WVC) over a subset of
reads $\lambda$,
\begin{equation}
U_{\lambda}(z)=\sum_{m}^{}s^{\lambda}_{m}z^{v^{\lambda}_{m}}.
\end{equation}
Let $W_{\Lambda}=\sum_{j \in \Lambda}w_j$ be the total weight of all the votes
belonging to the WVC, and let the total weight of the subsystem $\lambda$ be given
by $W_{\lambda}$. The weight not belonging to $\lambda$ can be expressed as
\begin{equation}
\sigma=\sum_{j \neq \lambda}w_j=W_{\Lambda}-W_{\lambda}.
\end{equation}
Note that if $W$ is the largest element of the vector $v^{\lambda}_{m}$ and
$v^{\lambda}_{m}(W)-v^{\lambda}_{m}(i)>\sigma$, then any element
$v^{\lambda}_{m}(i) \neq W$ can be set to zero since this does not affect the
probability of reliability even if all of the remaining votes are given to $i$.
Similarly, vectors satisfying $v^{\lambda}_{m}(W)-v^{\lambda}_{m}(1)>\sigma$,
$W \neq 1$ can be removed from further consideration for the same reason.
After these simplifications, the probability of correctly identifying the base is given by
\begin{equation}
P(\hat{b}_k = b_k) = r_k = \delta(U_{\Lambda},\hat{b}_k)=\sum_{\hat{b}_k (V^{\Lambda}_m)=b_k}
s^{\Lambda}_{m},
\label{rel}
\end{equation}
where $\Lambda =\{1,2,\ldots, c_k\}$.
The steps for computing the error probability of decision for a given set of
read positions are summarized below.
\begin{algorithm}
\caption{Computation of $P(\hat{b}_k =b)$}
\label{algo2}
\begin{algorithmic}
\STATE {\it 1. For each observations $y^{(i)}_k, 1\leq i \leq c_k$, define $U_{i}(Z)$ according
to (\ref{HP1}). }
\STATE {\it 2. Determine U-functions $U_{\Lambda}(z)$ for entire reads (in an arbitrary order) by
applying (\ref{twoterms1}) and (\ref{generalterm}) and collecting identical terms in the
intermediate U-function.}
\STATE {\it 3. Simplify expressions (zeroing or removing as discussed).}
\STATE {\it 4. Apply $\delta$-operator (\ref{rel}) to obtain the probability of correct
classification $r_k$ and the probability of error as $1-r_k$. }
\end{algorithmic}
\end{algorithm}

Having computed the probability of error $P(\hat{b}_k \ne b_k)$ for a fixed coverage
$c_k$ (where the MAP estimator forms $\hat{b}_k$ by combining information from
$c_k$ reads that cover the $k^{th}$ base), we can readily evaluate the probability of
error for a random coverage. In particular, the average assembly error probability
$P_{error}$ can be found by evaluating
\begin{equation}
P_{error}=\sum_{c_k} P(c=c_k) P(\overline{err}_{c}|c=c_k),
\end{equation}
where $\overline{err}_{c_k}$ denotes the error averaged over different read
positions given a fixed $c_k$. The probability distribution of $c$ is often assumed to be
Poisson \cite{kao11}, \cite{Waterman2} with some parameter $\lambda$. Assuming a
non-zero coverage, the mean coverage is given by $\bar{c}=\frac{\lambda}{1-\exp(-\lambda)}$
\cite{Waterman2}. With bases not covered by any reads we associate the probability of error of
$\frac{3}{4}$. Thus we can write the average error probability (conditioned on the coverage
depth being at least $1$) as
\begin{equation}
P_{error}=\frac{1}{1-\exp(-\lambda)}\sum_{c_k=1}^{\infty}\frac{\lambda^c\exp^{-\lambda}}{c!}P(\overline{err}_{c}|c=c_k).
\label{Perror2}
\end{equation}
Note that the fraction of bases not covered by any read is given by $\exp(-\lambda)$ and
thus the overall probability of error is given by $P_{total}=\frac{3}{4}\exp(-\lambda)+P_{error}$.
For distributions other than Poisson, e.g., empirical distributions inferred from data, one
can still apply the above approach to perform a semi-analytical evaluation of $P_{error}$.

\vspace{5mm}
\section{Experimental Results}
\label{sec:results}

In this section we present performance studies using both simulations and experimental
data sets. First, using realistic synthetic data, we compare the performance of the message
passing algorithm from Section II (Algorithm \ref{mpsq}), the binary message passing
algorithm and the power iteration algorithm. Moreover, we examine the convergence
properties of all these schemes and benchmark their accuracy by comparing it with the
genie-aided MAP estimation employed in the idealistic scenario where the exact error
probabilities of the reads are known. Then we proceed by testing the algorithms on the
experimental data we obtained by sequencing {\it E. Coli} and {\it N. Meningitides} using
the Illumina's HiSeq sequencing instrument that provides $100$-bp long reads. In particular,
we compare the performance of our developed reference-guided sequence assembly
algorithms with the commonly used sequencing data analysis tools including GATK and
SAMtools.

\subsection{Simulation data}

We simulated reference-guided sequence assembly of the genome of a strain of
{\it Neisseria Meningitidis}. The reference sequence is obtained from GenBank
(http://www.ncbi.nlm.nih.gov/nuccore) database and is $L=2,184,406$ bases long.
The reference is used to generate target sequences having $1\%$ variation rate.
We then uniformly select starting positions along the sequence and simulate short
reads of length $l=76$ (mimicking Illumina's Genome Analyzer II platform).
Sequencing errors in these reads are simulated according to the position-dependent
base calling error profile typical of this particular sequencing platform \cite{Vikalo12}.
The average error rate of the base calling procedure is $0.015$ (averaged over all reads
and bases in the reads). To construct the bipartite graphical model, we map the reads
to the reference sequence using an alignment algorithm based on the Burrows-Wheeler
transform \cite{Langmead09} and thus establish connections (i.e., edges) between the read nodes
and their aligned base nodes. The read nodes with multiple candidate mapping positions
are replicated (where each replica may be assigned different confidence score), and each
replica is connected to its corresponding set of base nodes. The bipartite graph with
binary base nodes introduced in Section~\ref{sec:binary} is constructed in the same way.
We apply both the message passing algorithms from Section~\ref{model_algorithm} and
Section~\ref{sec:binary} to infer the target sequence (note that since the algorithms are
randomly initialized, the stopping points and hence the resulting assembled sequences
may be different). We also form the binary data matrix representing all the short read data
and employ the power iteration method to infer the target genome
sequence. While the analysis in Section~\ref{sec:binary} gives a guarantee of convergence
of the power iteration algorithm, we found that its convergence is usually faster than the
theoretical bound. We set the stopping criterion for all these iterative learning methods as
$\sum |y_{j \rightarrow i}^{(k)}-y_{j \rightarrow i}^{(k-1)}|<\epsilon=0.01L$. It turns out
that both message passing algorithms need $\sim 30$ iterations to converge, while the
power iterations converge in $\sim 50$ iterations. We initialize all these algorithm by
generating $y_{j \rightarrow i}^{(0)}$ from Gaussian distribution $\mathcal{N}(1,1)$ and
uniform distribution $U(0,1)$ -- our extensive simulation studies indicate that different
initializations lead to the same error rate of the considered iterative schemes.

For a comparison, we also consider the plurality voting based decision scheme often
used in practice (see, e.g., \cite{kao11}). Here, multiple calls for a base in any
given position along the target sequence are consolidated by performing plurality voting.
Notice that, in both message passing and plurality voting, we assume the error profiles
of the reads (i.e., base calling error rates) are unknown. Plurality voting assumes all
reads has equal reliability while message passing scheme iteratively infers the reliability
of each read. We also consider probability of error of the MAP decision scheme in
Section~III which assumes perfect knowledge of the positions of reads along the
target sequence and exact information about position-dependent base calling errors
(both assumptions are unrealistic in practice). The error rates of these algorithms are
shown in Fig.~3 for various sequencing coverages (horizontal axis shows the average
coverage). As can be seen from Fig.~3, the message-passing scheme and the power
iteration algorithm outperform plurality voting. The binary message passing algorithm
has almost identical accuracy as power iterations, while being slightly worse than
Algorithm~1. Moreover, we see that the error rates of message passing are
close to the genie-aided MAP decision scheme.

 \begin{figure}[htbp]
   \centering
   \includegraphics[width=0.5\textwidth]{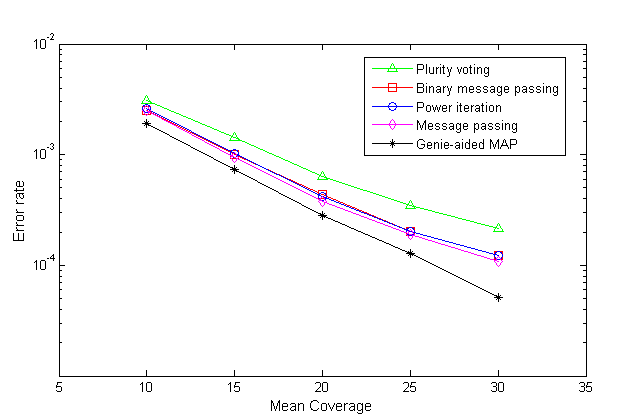}
   \caption{\it Error rates performance of the iterative learning schemes
(message passing, binary message passing, and power iterations) compared
with the plurality voting and genie-aided MAP schemes. The error rates of
iterative learning schemes and plurality voting are averaged over $20$
experiments.}
   \label{fig:err}
 \end{figure}

\subsection{Experimental data}

In addition to the simulation studies, we tested the performance of our proposed
iterative learning schemes for reference-guided sequence assembly using two experimental data
sets. In particular, we sequenced {\it Escherichia Coli} (from strain MG1655,
$4.64 \times 10^6$ bases long) and {\it Neisseria Meningitidis} (from strain FAM18,
having length $2.2 \times 10^6$) at the Center for Genomic
Sequencing and Analysis of the University of Texas at Austin. The data is obtained
using Illumina's HiSeq platform that provides $100$bp-long paired-end reads, and the
performance of our proposed methods are compared with that of the widely used
sequencing analysis packages SAMtools and GATK. Both SAMtools and GATK process
aligned next-generation sequencing data stored in SAM format, the alignment
file format provided by the majority of frequently used alignment tools (e.g.,
BWA). These files contain the aligned reads, their positions and the quality scores
of the bases. SAMtools calculates empirical quality scores from the alignment
information and uses them to recalibrate the raw quality scores provided by the
sequencing platform. The assembled sequence is formed using the aligned bases
weighted by these new quality scores. In addition to the quality score recalibration,
GATK also performs a local realignment procedure to correct misaligned reads,
especially from the target genome region containing indels compared to the reference
genome. After performing sequence assembly using quality score information,
these software packages can also perform downstream single nucleotide polymorphism
(SNP) detection, while GATK also incorporates a machine learning tool to separate true
variation from sequencing platform artifacts.

The two genomes are sequenced using $67\%$ of an HiSeq platform lane having
approximately $30 \times 10^6$ reads, resulting in the coverage greater than $200$.
This enables accurate inference of the true {\it E. Coli} and {\it N. Meningitides} sequences
using any of the techniques discussed in the paper, providing us with the ground truth.
To determine the accuracy of our proposed schemes in realistic scenarios where the
coverage is limited, we uniformly
subsample the data to emulate low coverage situations. The resulting error rates are
shown in Table~I. As can be seen there, the developed message passing schemes
outperform both SAMtools and GATK in terms of the accuracy. The number of iterations
for each message passing scheme was set to $30$, which at coverage $c=20$ resulted
in the average CPU runtimes of $65$ and $37$ minutes for processing {\it E. Coli} and
{\it N. Meningitidis} data sets, respectively (the algorithms were coded in C++, run on a 3.07G Hz
single core machine). The corresponding runtimes for SAMtools are $50$ and $28$ minutes,
and for GATK $53$ and $30$ minutes. As seen from the table, increasing the coverage
can dramatically improve accuracy of the assembly -- recall the discussion from Section V
where we showed that the
probability of error of the genie-aided MAP estimator decreases exponentially with the
coverage. However, increasing coverage also increases the cost of the sequencing project.

\begin{table}
\begin{center}
\label{tab:error}
\begin{tabular}{c c c c c}
  \hline
  Sequence & \multicolumn{4}{c}{Number of errors}   \\
  \cline{2-5}
   and Coverage& MP & BMP & SAMtools & GATK \\
  \hline
  E coli & & & & \\
  \hline
  15 & $3484\pm 88$ & $3507 \pm 76$ & $3655\pm66$& $3598\pm72$\\
  20 & $2566\pm 66$ & $2599\pm54$ & $2677\pm71$ & $2634\pm53$\\
  25 & $1243\pm31$ & $1256\pm44$ & $1298\pm41$ & $1283\pm 55$\\
  30 & $763\pm20$ & $781\pm23$ & $811\pm23$ & $798\pm18$\\
  \hline
  N. Meningitidis & & & & \\
  \hline
  15 & $2168\pm 35 $& $2231\pm43 $& $2404\pm 37$ & $2358\pm30$\\
  20 & $1201\pm 29$ & $1299\pm 26$ & $1388\pm 30$ &$1379\pm20$\\
  25 & $899\pm 16$& $913\pm 20$& $933\pm24$&$921\pm19$ \\
  30 & $658\pm11$& $669\pm11$& $681\pm9$&$680\pm15$ \\
  \hline
\end{tabular}

\caption{\it Performance of the message passing algorithm (MP), binary message
passing algorithm (BMP), SAMtools and GATK on E. coli and N. Meningitidis
sequencing data with various coverages. The average number of decision errors
and the corresponding standard deviation are computed over $30$ runs.}
\end{center}
\end{table}

Note that the sequenced genome might contain insertions as compared to the reference
or, equivalently, the reference sequence contains gaps.
This structural variation can be detected in the alignment stage by using paired-end reads
\cite{Dnastar}, \cite{Rausch09}. The paired-end reads have a known range of lengths of inserts
between the reads in a pair. The gaps in the reference can be detected by relying on a multi-read
alignment of the pairs of reads and comparing the aligned positions with the insert lengths. We used the
scheme in \cite{Rausch09} to perform the alignment of our {\it E. Coli} data set
and detected $34$ gaps in the reference. We includes the gap positions as additional base nodes in
our graphical model and uses our Algorithm 1 to identify the order of nucleotides in the gaps.
As a result, $31$ out of $34$ gaps were reconstructed (i.e., closed).

%

\vspace{5mm}
\section{Summary and Conclusion}
\label{sec:conclusion}

We studied reference-guided sequence assembly from short reads generated by
next-generation sequencing technologies, specifically focusing on the problem of
obtaining the target genome sequence from potentially erroneous and misaligned
reads. We cast the problem as the inference of the target sequence on an
appropriately defined bipartite graph
and proposed iterative learning algorithms for solving it. In particular, we developed
message passing algorithms that rely on both binary as well as representation of
nucleotide bases by $4$-dimensional vectors. It was shown that the derived
message passing algorithm (in particular, Algorithm~1 in Section~II) can be
interpreted as the standard belief propagation under a certain prior. In addition,
the problem was rephrased
so that the power iteration algorithm, employed to find the leading singular vector
of a matrix collecting all short reads, results in a good approximation of the
target sequence. Convergence of power iterations is guaranteed, while the
convergence of message passing algorithms is studied empirically. Unlike existing
methods, the proposed algorithms find the desired sequence without using
reliability information (i.e., quality scores) of the short reads -- in fact, message
passing algorithms infer the aforementioned quality score information.

To assess achievable accuracy of the proposed iterative learning techniques, we analyzed
the probability of error of a genie-aided maximum a posteriori decision scheme
in the idealized scenario where the base calling error rates and read mapping locations
are known perfectly. It was shown empirically that the iterative learning schemes
perform close to the genie-aided estimation scheme, and that they outperform
state-of-the-art software packages for downstream processing of sequencing data.

\section*{Acknowledgment}

This work is funded by the National Institute of Health under grant 1R21HG006171-01.
We thank Dr. Devavrat Shah for pointing out the reference \cite{Shah11} and useful
discussions.


\end{document}